\documentstyle[12pt]{article}          

\def	\eqnum		#1{(\ref{#1})}       
\def	\scite		#1{$^{\cite{#1}}$}     
	\newdimen\eqskip
	\newdimen\txtskip
	\baselineskip=\txtskip
	
\begin{document}

\vspace*{-1in}
\begin{flushright}
Fermilab-Conf-95/308-T\\
\end{flushright}

\vspace*{0.75in}
\begin{center}
       	{ \Large  \bf \sc
THE SOLAR NEUTRINO PUZZLE;\\ UPDATE
	}
\footnote{Invited talk at the XV International Conference
of Physics in Collision, Krakow, Poland, June 8 - 10, 1995.}

\vskip 1.0in

{\it STEPHEN PARKE }\\[0.05in]
{\small
parke @ fnal.gov \\[0.1in]
Department of Theoretical Physics\\
Fermi National Accelerator Laboratory\\
P.O. Box 500, Batavia, IL 60510, U.S.A.} \\[0.2in]
\end{center}

\vskip 1.5in
\begin{abstract}
In this talk I will summarize the latest experimental results
from the four solar neutrino experiments and discuss what this means for
the flux of $^7Be$ and $^8B$ neutrinos.
The implications for the solar models including the new versions with helium
and heavy element diffusion will also be addressed.
The exciting and important calibration results of the Gallex collaboration
will be presented as well as the outlook for the next generation
of solar neutrino experiments.
\end{abstract}

\def    \He          {\mbox{$^{4}He$}}
\def    \Be		{\mbox{$^{7}Be$}}
\def    \Bo		{\mbox{$^{8}B$}}
\def    \nupp		{\mbox{$\nu_{e}^{pp}$}}
\def    \nupep		{\mbox{$\nu_{e}^{pep}$}}
\def    \nuBe		{\mbox{$\nu_{e}^{^7 {Be}}$}}
\def    \nuBo		{\mbox{$\nu_{e}^{^8 B}$}}
\def    \f		{\mbox{$\phi$}}
\def    \fpp		{\mbox{$\f^{pp}$}}
\def    \fpep		{\mbox{$\f^{pep}$}}
\def    \fCNO		{\mbox{$\f^{CNO}$}}
\def    \fBe		{\mbox{$\f^{^{7}{Be}}$}}
\def    \fBo		{\mbox{$\f^{^{8}B}$}}
\def    \nfpp		{\mbox{${\Phi}^{pp}$}}
\def    \nfpep		{\mbox{${\Phi}^{pep}$}}
\def    \nfBe		{\mbox{${\Phi}^{^{7}{Be}}$}}
\def    \nfBo		{\mbox{${\Phi}^{^{8}B}$}}
\def    \SeC		{\mbox{$S^{ex}_{Cl}$}}
\def    \SeH		{\mbox{$S^{ex}_{H_2O}$}}
\def    \SeD		{\mbox{$S^{ex}_{Home}$}}
\def    \SeK		{\mbox{$S^{ex}_{Kam}$}}
\def    \StC		{\mbox{$S^{th}_{Cl}$}}
\def    \StH		{\mbox{$S^{th}_{H_2O}$}}
\def    \StG		{\mbox{$S^{th}_{Ga}$}}
\def    \SeG		{\mbox{$S^{ex}_{Ga}$}}
\def    \SeSe		{\mbox{$S^{ex}_{Sage}$}}
\def    \SeGx		{\mbox{$S^{ex}_{Gallex}$}}
\def    \X		{\mbox{$\chi^2$}}
\def    \BP		{\mbox{Bahcall \& Pinsonneault}}
\def    \TCL		{\mbox{Turck-Chi\`{e}ze \& Lopes}}

\def   	\bea            {\begin{eqnarray}}
\def   	\eea            {\end{eqnarray}}
\def   	\beq            {\begin{equation}}
\def   	\eeq            {\end{equation}}
\def 	\nn		{\nonumber}

\newpage
\section*{Current Experimental Situation}
Over the last year new results from the four solar neutrino
experiments have been reported.
The results for Homestake\scite{erc}, Kamiokande\scite{erh},
Gallex\scite{erg} and SAGE\scite{ers} are
\bea
        \SeD & = & 2.55 \pm 0.17 \pm 0.18  ~~~SNU               \nn \\
        \SeK & = &  0.51 \pm 0.04 \pm 0.06 ~~~\Phi^{^8B}_{BP}   \nn  \\
        \SeGx & = & 79 \pm 10 \pm 6  ~~~SNU                     \nn \\
        \SeSe & = & 69 \pm 11~^{+5}_{-7}  ~~~SNU                \nn
\eea
where the first uncertainty is statistical and second systematic.
To form a combined result for gallium, the mean and statistical errors
for SAGE and Gallex were combined in the standard way
but a common systematic error of 6 SNU was used.
Then the statistical and systematic errors
are combined in quadrature for each experimental result giving
\bea
        \label{ec}
       \SeC  & = & 2.55 \pm 0.25 ~~~SNU    \\
        \label{eh}
        \SeH & = & 0.51 \pm 0.072 ~~~\Phi^{^8B}_{BP}    \\
        \label{eg}
        \SeG   & = &   74 \pm 9.5  ~~~SNU.
\eea

To compare these experimental results with those from the standard solar models
it is convenient to use one of the models as a reference model.
I will use the 1992
solar model of Bahcall and Pinsonneault \scite{BP} as this reference
solar model where the central values of the important
solar neutrino fluxes are
\bea
\Phi^{pp}_{BP}  & = &  6.0 \times 10^{10} ~cm^{-2} sec^{-1} \nn \\
\Phi^{^7Be}_{BP} & = &  4.9 \times 10^{9} ~cm^{-2} sec^{-1} \nn \\
\Phi^{^8B}_{BP} & = & 5.7 \times 10^{6} ~cm^{-2} sec^{-1}. \nn
\eea
It is useful to normalize all solar neutrino fluxes to this model,
by defining the renormalized neutrino fluxes as
\beq
\f^i ~=~ \Phi^i / \Phi^i_{BP}.
\eeq
In these normalized flux units the solar luminosity constraint
is simply
\bea
        \label{lum}
        1 & = & 0.913~\fpp ~+~ 0.071~\fBe ~+~ 4 \times 10^{-5}~\fBo
\eea
This will be used to determine \fpp ~in terms of \fBe.
Then the contribution of the $\nupp, ~\nuBe$ and $\nuBo$
to the chlorine, water and gallium solar neutrino experiments is
\bea
        \label{tc}
        \StC & = & 6.2~\fBo ~+~ 1.2~\fBe  ~~~SNU  \\
        \label{th}
        \StH & = & \fBo  ~~~\Phi^{^8B}_{BP}      \\
        \label{tg0}
        \StG & = & 14~\fBo ~+~ 36~\fBe ~+~ 71~\fpp  ~~~SNU.
\eea
The coefficients in eq.\eqnum{tc}-\eqnum{tg0}
are determined using the assumptions that the state of the neutrinos
is unaffected by the passage from the solar core to the terrestrial detectors,
i.e. there is no change in the flavor, helicity or energy spectrum,
and that the neutrino interaction cross sections used are corrected.
The uncertainty on these cross sections is estimated to be a few per cent.

\begin{figure}[t]
\vspace{8cm}
\includegraphics{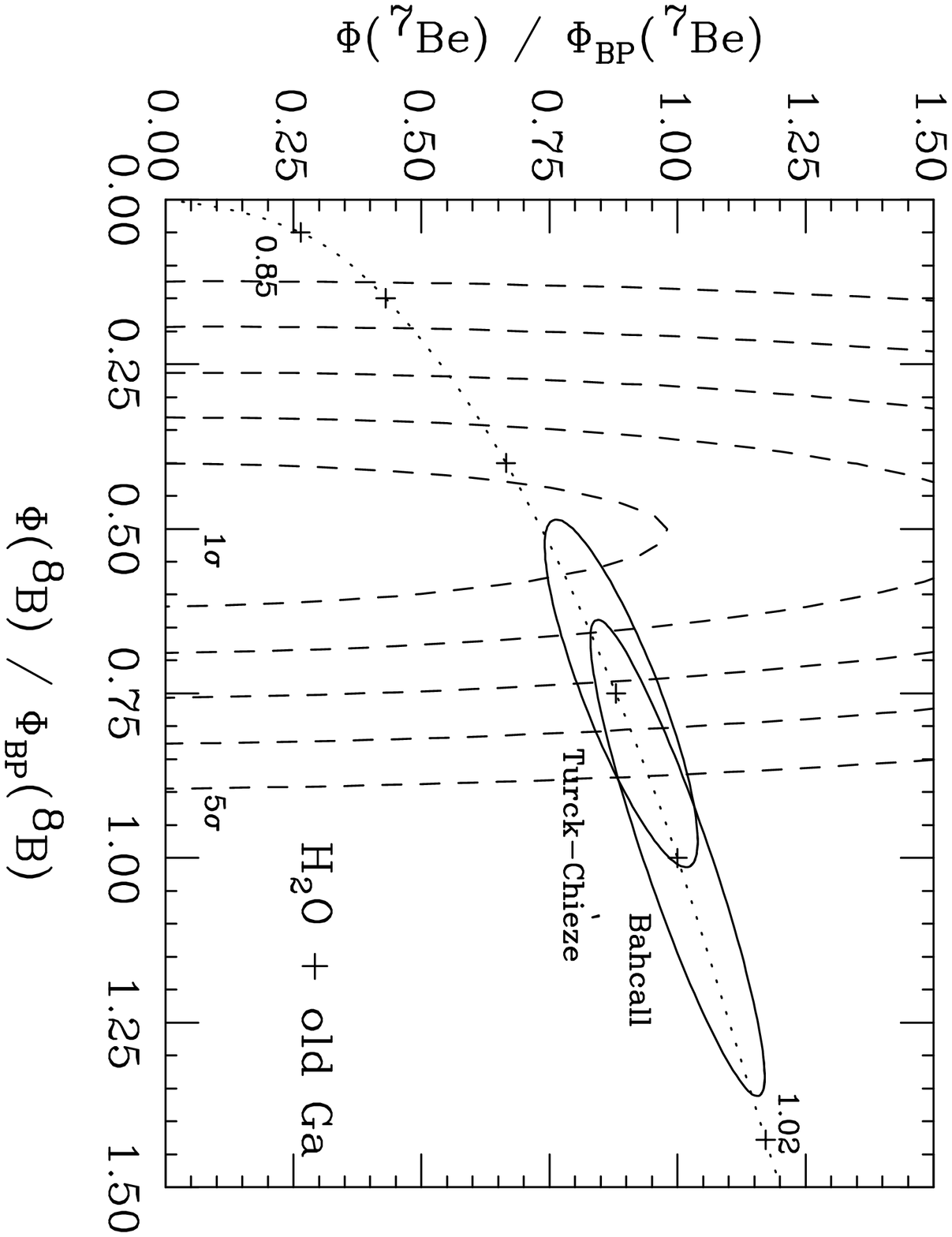}
\includegraphics{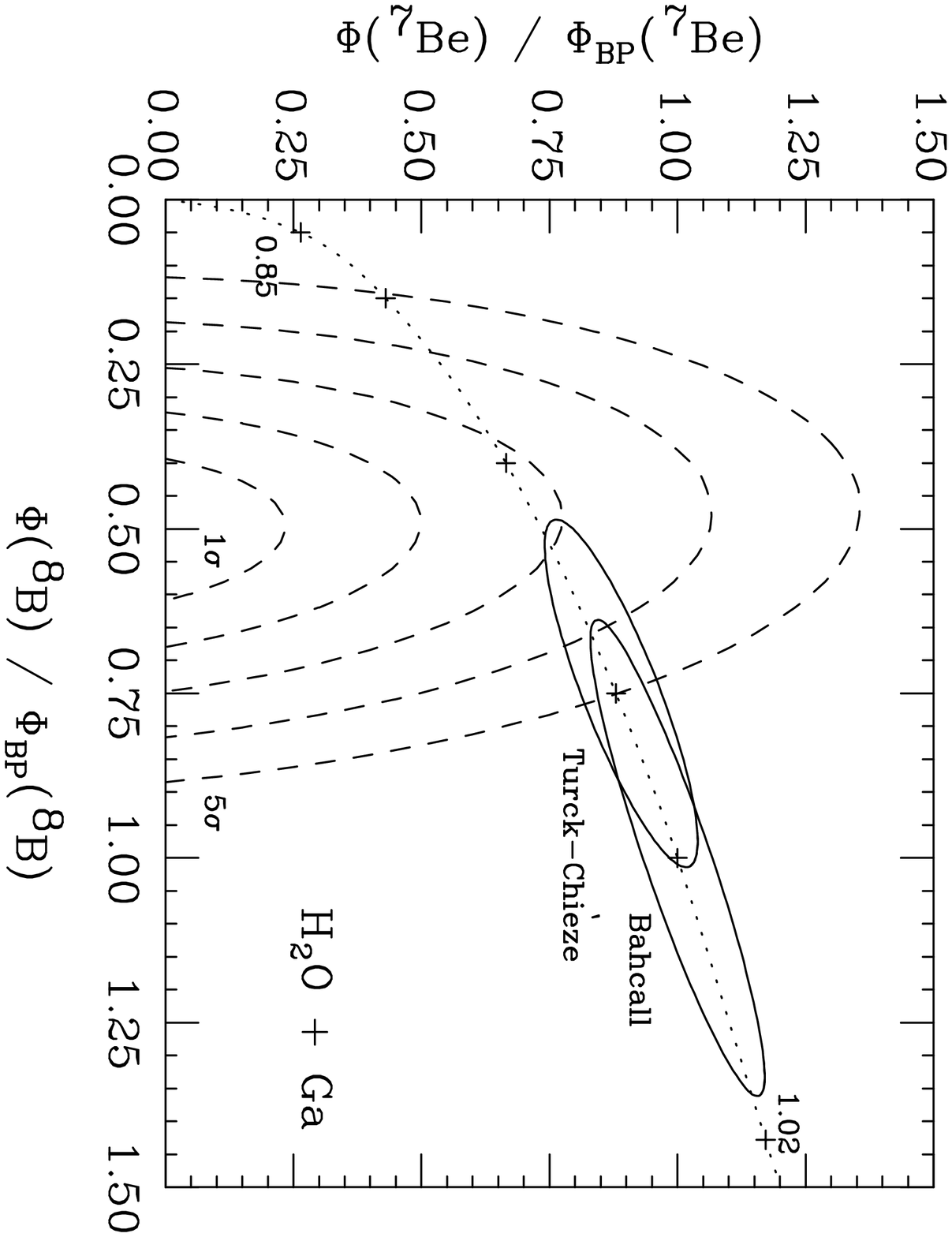}
\vspace{0.5cm}\hspace{1cm}(a)\hspace{8cm}(b)\vspace{1cm}
\vspace{-1.0cm}
\caption[]{(a)
The \fBe~ verses \fBo~ plane using the results from Kamiokande
and the old Gallex results.
The dashed curves are the 1$\sigma$ to
5$\sigma$ contours for the $\chi^2$ variable.
The solid ellipses are the predictions of the solar
models of Bahcall \& Pinsonneault 1992 and Turck-Chi\`{e}ze \& Lopes.
The dotted line is the curve $\fBe = (\fBo)^{8/18}$
and the crosses on this line corresponding to solar core temperature of
(0.85, 0.90, 0.95, 0.984, 1.00, 1.02) times the
core temperature of the Bahcall \& Pinsonneault's model.
(b) Same as (a) but using the latest combined results from
Gallex and SAGE as well as Kamiokande.}
\label{kamgal}
\end{figure}

Using the luminosity constraint to eliminated the $\nupp~$ flux, the
contribution to the gallium experiments can be written as
\bea
        \label{tg}
        \StG & = & 14~\fBo ~+~ 30~\fBe ~+~ 78   ~~~SNU.
\eea
The additional contributions from other specifies
of neutrinos is less than 10\%
in the standard solar models \scite{CNO}.

\begin{figure}[bht]
\vspace{11cm}
\includegraphics{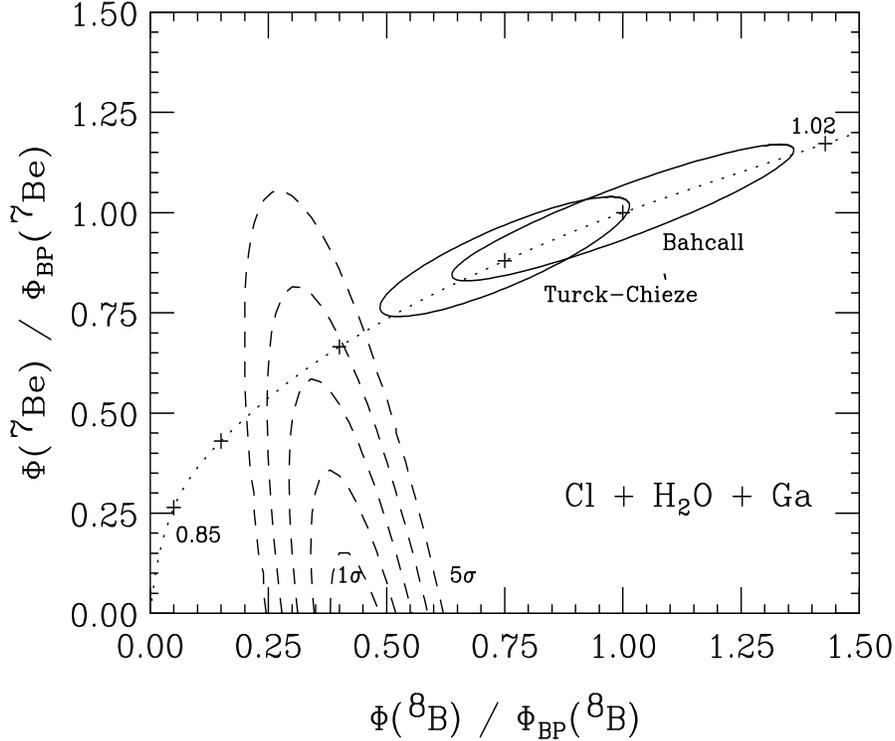}
\caption[]{Similar to Fig. 1 but using the latest experimental results
from all four solar neutrino experiments.}
\label{models}
\end{figure}

The experimental results, \eqnum{ec}, \eqnum{eh} and \eqnum{eg},
are now used to fit the two parameters, \fBe~ and \fBo~,
of the model, eq.\eqnum{tc}, \eqnum{th} and \eqnum{tg}.
The \X~ variable was calculated for the four cases;
all three results together and the three ways of choosing two out of three.
Since the minimum value of \X~ occurs at negative values of \fBe~
for all four cases, the constraint
\bea
\fBe & \ge & 0
\eea
was imposed \scite{PDG}. Fig. 1 shows the difference in the exclusion
using only the Kamiokande and new Gallium (Gallex and SAGE combined)
as compared to the previous result from Gallex ($83 ~\pm~ 21 ~SNU$).
The total theoretical range for the solar models of
Bahcall and Pinsonneault (1992) \scite{BP}
and Turck-Chi\`{e}ze \& Lopes \scite{TCL}
are also shown in this figure by the labelled ellipses.
Fig. 2 shows the exclusion using all of the latest solar neutrino
experimental results. This argument was first presented by the authors of
ref. \cite{first} and updated by the authors of ref. \cite{update}.

Bahcall \scite{BBe} has argued that by using the Kamiokande measurement
to determine $\fBo$ then the Chlorine experiment puts an upper limit
on the $\fBe$ at the 95\% C.L. equal to 0.41, that is,
\bea
\Phi^{^7Be} & < &  2.0 \times 10^{9} ~cm^{-2} sec^{-1}.
\eea
Similarly, he has used the Gallium plus Kamiokande plus Luminosity
constraint to show that $\fBe ~<~ 0.53$ at 95\% C.L. that is,
\bea
\Phi^{^7Be} & < &  2.6 \times 10^{9} ~cm^{-2} sec^{-1}.
\eea

These results suggest that $\fBe ~<~ \fBo $. Remember however that
both the $^7Be$ and $^8B$ neutrinos are produced from the same
parent in the sun, that is, $^7Be$ via electron and proton interactions
respectively. Also the $^8B$ neutrinos are more sensitive to changes
in the solar core temperature, $T_c$, than the $^7Be$ neutrinos,
$T_c^{18}$ verses $T_c^8$
respectively.
Therefore it is very difficult to arrange $\fBe ~<~ \fBo ~<~ 1$
in standard solar models.

\begin{figure}[bht]
\vspace{9cm}
\caption[]{Characteristics of the decay of $^{51}Cr$. The ''751 keV'' line
combines the 746 and 751 keV lines and ''431 keV'' line combines
the 426 and 431 keV lines.}
\label{chromium}
\end{figure}

\section*{Calibration of the Gallium Experiment}

{}From June to October 1994 the Gallex detector \scite{gallexcal}
was exposed to a
$61.9 ~\pm~ 1.2 ~PBq$ neutrino source which emits neutrinos in
electron capture in $^{51}Cr$, see Fig. 3.
This source was made by bombarding
enriched chromium in a nuclear reactor.
The initial source activity produced a flux of neutrinos
at the detector which was approximately 15 times the solar neutrino flux.
This collaboration used three different methods to measure the initial
source strength; neutron flux capture calculation, calorimetry and by
measuring the 320 keV gamma ray emitted from small samples.
The average of these
measurements was used to compare with the
strength obtained from observing the
neutrino capture in the Gallex detector of $64.1 ~\pm~ 6.6 ~\pm~ 3.3 ~PBq$,
see Fig. 4.
\begin{figure}[htb]
\vspace{10cm}
\caption[]{
Number of $^{71}Ge$ atoms produced per day during the course of the
source experiment (first 7 runs only). The points for each run
are plotted at the beginning of each exposure, with the horizontal
lines showing the duration of the exposures. The predicted curve
(dotted line), which decreases with the known half-life of $^{51}Cr$,
is based on the relationship between the directly measured source strength
and the 0.189 $^{71}Ge$ production rate per day. The curve also
includes the constant 0.78/day production rate due to solar
neutrinos and side reactions (dashed line).
}
\label{allfour}
\end{figure}

The ratio of the source activity as measured by Gallex to that obtained from
the other methods was
\beq
1.04 ~ \pm~ 0.12.
\eeq
This result validates the radiochemical methods of the Gallex experiment and
since 90\% of the neutrinos from the $^{51}Cr$ source have an energy close to
the energy of the $^7Be$ neutrinos the Gallex experiment is fully efficient
at detecting neutrinos of this energy. This is a very important
milestone for solar neutrino experiments.

In the autumn of 1995 the $^{50}Cr$ will be re-irradiated and the
calibration will be repeated.
SAGE is  also performing a calibration test and counting of samples from
this test will continue throughout the summer of 1995.

\section*{Improved Solar Models}

\begin{figure}[h]
\vspace{10cm}
\includegraphics{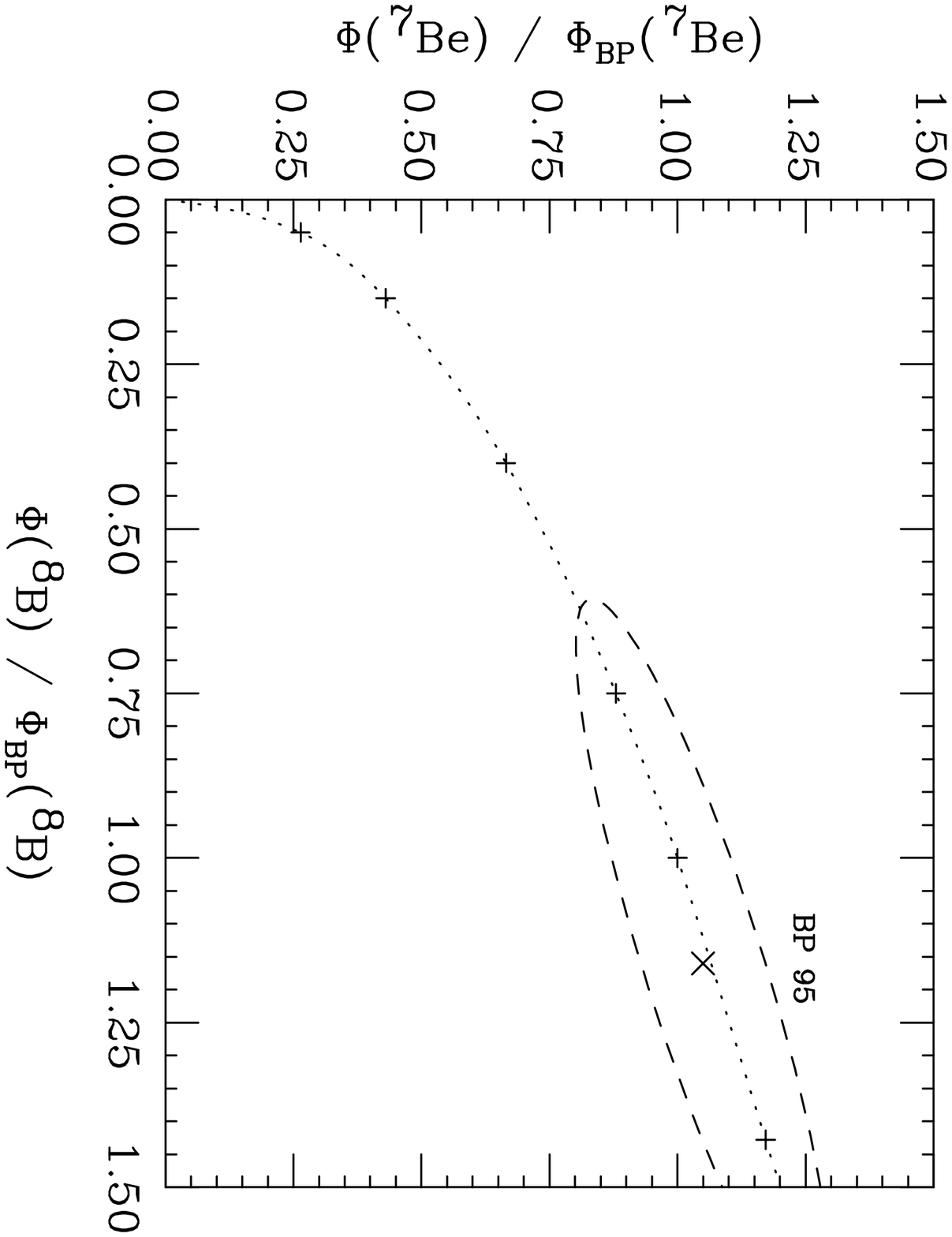}
\includegraphics{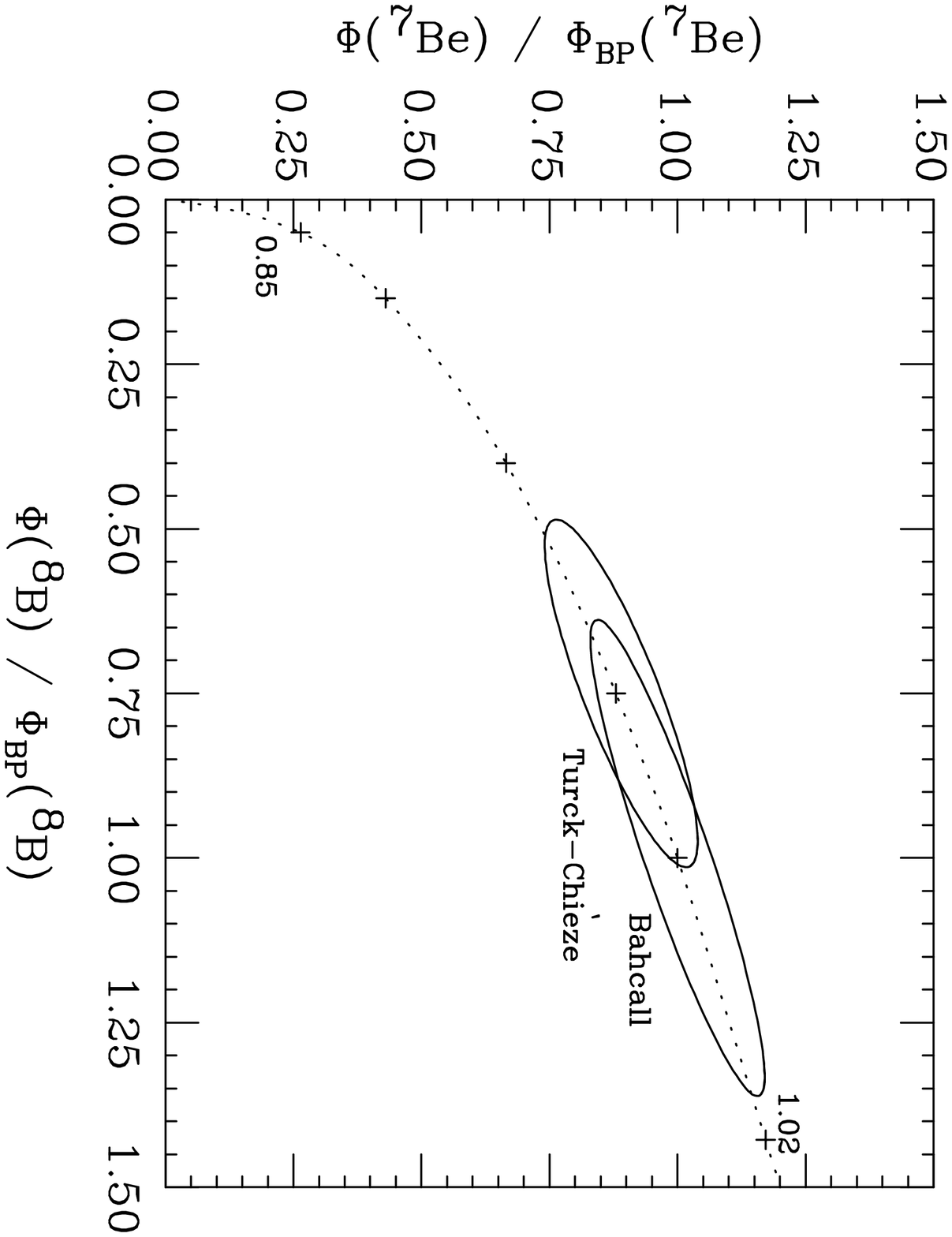}
\caption[]{
Comparison between the Bahcall \& Pinsonneault '95 solar model
(dashed ellipse) and solar models of
Turck-Chi\`{e}ze \& Lopes and
Bahcall \& Pinsonneault '92 (solid ellipses). The cross in the center
of the dashed ellipse is the central value for the BP95 model.
}
\label{diffusion}
\end{figure}

The inclusion of helium and heavy element
diffusion has improved the consistency
of the solar models by
Proffit \scite{proffit},
Kovetz and Shaviv \scite{kovshav} and
Bahcall and Pinsonneault \scite{bahpin}
with helioseismology.
The important parameters
are the surface abundance of helium,
\beq
Y_S ~=~ 0.242 \pm 0.003 \nn
\eeq
and the depth of the convective zone,
\beq
R_{CZ} ~=~ 0.713 ~\pm~ 0.003~R_{\bigodot}. \nn
\eeq
Bahcall and Pinsonneault '95 ('92)
models give the surface abundance of helium at 0.247 (0.273) and
the fractional depth of the convective zone as 0.712 (0.707) respectively.
Clearly the inclusion of diffusion improves the
consistency in these parameters.

\begin{figure}[hbt]
\vspace{9cm}
\caption[]{Allowed parameter space for the ``Just-so'' oscillation solution
to the solar neutrino puzzle by Krastev and Petcov. (a) includes and (b)
does not include the theoretical uncertainties in the analysis.
}
\label{justso}
\end{figure}

However the flux of both $^7Be$ and $^8B$ neutrinos increases
compared to their '92 model
\bea
	\Phi_{^7Be} & = & 5.1~(1\pm0.06) \times 10^9 ~cm^{-2} s^{-1} \\
	\Phi_{^8B} & = & 6.62~(1\pm0.16) \times 10^6 ~cm^{-2} s^{-1}
\eea
This increase in fluxes leads to an increase in the expected
Chlorine and Gallium counting
rates to $9.3 \pm 1.3$ SNU and $137 \pm 8$ SNU as well
as an increase in the flux for the Kamiokande experiment.
The effect of these increased fluxes on our comparison of theory verses
experiment for the solar neutrino flux is shown in Fig. 5.
Clearly these new models do not help resolve the discrepancy between
the solar models and the solar neutrino experiments.

\section*{Neutrino Oscillation Solutions}
The latest iso-SNU contour plots by Krastev and Petcov \scite{krastpet}
for the ``Just So'' solution is given in Fig. 6 and for
the MSW solution by Hata and Langacker \scite{hatalang} in Fig. 7.

\begin{figure}[h]
\vspace{9cm}
\caption[]{The updated result of the combined MSW analysis
assuming the Bahcall-Pinsonneault '92 model, see Hata and Langacker.
}
\label{msw}
\end{figure}

\section*{Next Generation Experiments}
{\it SuperKamiokande} ~is 10 times larger than Kamiokande III with
a fiducial volume of 22 ktons and 11,000 20'' PMTs. This detector will
observe about $10^4$ solar neutrino events per year in the
neutrino-electron elastic scattering mode,
\bea
\nu_x ~+~ e & \rightarrow & \nu_x ~+~ e
\eea
and hopes to observe
distortions in the solar neutrino energy spectrum
after about two years of running, see fig. 8. This mode is primarily
sensitive to electron-neutrinos.

\begin{figure}[t]
\vspace{9cm}
\caption[]{The Super-Kamiokande electron spectrum expected for the
adiabatic and large-angle solutions, from Hata and Langacker.
}
\label{superk}
\end{figure}
\begin{figure}[bht]
\vspace{9cm}
\caption[]{The SNO charged current spectrum expected for the adiabatic
and large-angle solutions, from Hata and Langacker.
}
\label{sno}
\end{figure}

As of the time of this conference many of the PMTs had been checked
and pre-assembled. The start of installation was expected in June 1995
and completion in March 1996 with physics scheduled for April 1996.

{\it Sudbury Neutrino Observatory} ~consists of 1000 tons of heavy water
surrounded by a light water shield. This detector will be able to
observe solar neutrinos in three modes,
\bea
\nu_x ~+~ e & \rightarrow & \nu_x ~+~ e \\
\nu_e ~+~ d & \rightarrow & e ~+~ p ~+~ p \\
\nu_x ~+~ d & \rightarrow & \nu_x ~+~ p ~+~ n
\eea
where $x$ represents $e, ~ \mu$ or $\tau$. The expected rates for these
reactions is $10^3$, $10^4$ and $3 \times 10^3$ events per year.
The second of these modes will be able to measure the solar electron neutrino
spectrum, see Fig.   9, whereas the last reaction will measure the total
solar neutrino flux regardless of the neutrinos flavor.
At the time of this conference construction of this detector was
proceeding according to schedule with completion set for spring/summer 1996.

\begin{figure}[hbt]
\vspace{9cm}
\caption[]{Seasonal variation of the $^7Be$ signal via ``just-so'' vacuum
oscillations. Shown for comparison is also the
$1/R^2$ effect arising from the earth's motion only, from Borexino proposal.
}
\label{borexino}
\end{figure}

{\it Borexino} ~detector consists of 100 tons of liquid scintillator
with a very low threshold 0.25 MeV. Again this detector is sensitive to
\bea
\nu_x ~+~ e & \rightarrow & \nu_x ~+~ e
\eea
\clearpage
\noindent
but with such a low threshold this detector will be
sensitive to $^7Be$ neutrinos. If the standard solar model fluxes is
correct this detector can expect 20,000 events per year. For the MSW
solution to the solar puzzle the rate will be much less. Because of
the large event rate this detector will be able to see the $1/R^2$
variation in the solar neutrino flux.
Also this detector is very sensitive to
neutrino oscillations in the ``Just-so'' scenario, see Fig. 10.
As of May 1995 this collaboration
had demonstrated that they can achieve the purity levels required
to set a 0.4 MeV threshold in a 6 ton prototype.

\section*{Conclusions}
The calibration of the Gallex detector is a very important milestone
for the field of solar neutrinos giving us confidence in all of the
radio-chemical solar neutrino experiments.
With the turning on of SuperKamiokande and the Sudbury Neutrino Observatory
next year and Borexino a few years later this is an exciting time for
the field of solar neutrinos. We will learn whether or not the solar neutrino
puzzle is new, exciting neutrino physics or some problem with our understanding
of the solar interior. These experiments must resolve this issue as
soon as possible.


\newpage
\begin{thebibliography}{99}
\parskip=0pt

\def    \astro     #1#2#3{{ Astrophys. J.} {\bf #1},  #3, (#2)}
\def    \nuke   #1#2#3{{ Nucl. Phys.} {\bf B#1},  #3, (#2)}
\def    \pl     #1#2#3{{ Phys. Lett.} {\bf #1B},  #3, (#2)}
\def    \prl    #1#2#3{{ Phys. Rev. Lett.} {\bf #1},  #3, (#2)}
\def    \pr     #1#2#3{{ Phys. Rev.} {\bf #1},  #3, (#2)}
\def    \prd    #1#2#3{{ Phys. Rev.} {\bf D#1},  #3, (#2)}
\def    \prep   #1#2#3{{ Phys. Rep.} {\bf #1},  #3, (#2)}
\def    \rmp    #1#2#3{{ Rev. Mod. Phys.} {\bf #1},  #3, (#2)}
\def    \zeit   #1#2#3{{ Z. Phys.} {\bf C#1},  #3, (#2)}
\def    \cmp    #1#2#3{{ Comm. Math. Phys.} {\bf #1},  #3, (#2)}
\def    \ibid   #1#2#3{{\it ibid.} {\bf #1}, #3, (#2)}
\def    \jetp   #1#2#3{{ JETP Lett.} {\bf #1},  #3, (#2)}
\def    \sovnuke #1#2#3{{ Sov. J. Nucl. Phys.} {\bf #1},  #3, (#2)}

\bibitem{erc}
K.~Lande for the Homestake Collaboration, Neutrino 94, Israel, June 1994.
\bibitem{erh}
Y.~Suzuki for the Kamiokande III Collaboration, Neutrino 94.
\bibitem{erg}
T.~Kirsten for the Gallex Collaboration, Neutrino 94.
\bibitem{ers}
J.~Nico for the SAGE Collaboration, 27th ICHEP, United Kingdom, July 1994.
\bibitem{BP}
J.~N.~Bahcall and M.~H.~Pinsonneault, \rmp{64}{1992}{885}.
\bibitem{CNO}
Even if the CNO and pep neutrinos where included in the luminosity
constraint, the coefficients of \fCNO~ and \fpep~ would be positive
in eq. \eqnum{tg} so that any contribution from these sources would
strengthen the argument.
\bibitem{PDG}
This is the procedure recommended by the Particle Data Group,
\prd{45}{1992}{Part II} and gives a more conservative result
than not imposing this constraint.
\bibitem{TCL}
S.~Turck-Chi\`{e}ze and I.~Lopes, \astro{408}{1993}{347}.
\bibitem{first}
M.~Spiro and D.~Vignaud, \pl{242}{1990}{279},\\
V.~Castellani, S.~Degl'Immocenti and G.~Florentini,\\
Astron. Astrophys. {\bf 271}, 601, (1993),\\
N.~Hata, S.~Bludman and P.~Langacker, \prd{49}{1994}{3622}.
\bibitem{update}
S.~Parke, \prl{74}{1995}{839},\\
S.~Degl'Innocent, G.~Fiorentini and M.~Lissia, INFN-FE-10-94\\
N.~Hata and P.~Langacker, \prd{52}{1995}{420}.
\bibitem{BBe}
J.~N.~Bahcall, \pl{338}{1994}{276}.
\bibitem{gallexcal}
Gallex Collaboration, \pl{342}{1995}{440}.
\bibitem{proffit}
C.~Proffit, \astro{425}{1994}{849}.
\bibitem{kovshav}
A.~Kovetz and G.~Shaviv, \astro{426}{1994}{787}.
\bibitem{bahpin}
J.~N.~Bahcall and M.~H.~Pinsonneault, IASSNS-AST 95/24.
\bibitem{krastpet}
P. Krastev and Petcov, \prl{72}{1994}{1960}.
\bibitem{hatalang}
N.~Hata and P.~Langkacker, \prd{50}{1994}{632}


\end{thebibliography}
\end{document}